\documentstyle[12pt,fleqn,epsf]{ioplppt} % use this for preprint style

\def\a{\alpha}\def\d{\delta}
\def\i{\iota}
\def\q{\psi}\def\r{\rho}\def\s{\sigma}
\def\y{\eta}

\def\D{\Delta}\def\F{\Phi}
\def\O{\Omega}\def\P{\Pi}\def\S{\Sigma}

\def\mo{{-1}}

\def\ex{{\rm e}}\def\ds{ds^2=}
\def\fe{field equations }\def\bh{black hole }

\def\bg{background }
\def\ssy{spherically symmetric }

\def\GB{Gauss-Bonnet }

\def\be{\begin{equation}}\def\ee{\end{equation}}
\def\bea{\begin{eqnarray}}\def\eea{\end{eqnarray}}
\def\le{\lefteqn}\def\nn{\nonumber}
\def\lapp{[(r-2m)}
\def\alp{\alpha '}\def\GB{Gauss-Bonnet }\def\Sch{Schwarzschild }
\def\ri{{\cal R}}\def\gb{{\cal S}}\def\ef{{\rm e}^{-2\Phi}}
\def\es{{\rm e}^{-2\Sigma}} \def\efs{({\rm e}^{-2\F}+\d{\rm e}^{-2\S})}
\def\fo{\Phi_1}\def\ft{\Phi_2}\def\ro{\rho_1}\def\rt{\rho_2}\def\so{\S_1}
\def\st{\S_2}\def\po{\psi_1}\def\pt{\psi_2}\def\fsq{(\F'^2+\S'^2)}
\def\lap{[r(r-2m)}\def\bc{boundary conditions }\def\af{asymptotically flat }
\def\orq{o\left({1\over r^2}\right)}
\def\mt{{-2}}\def\mn{{mn}}

\begin{document}

\title{Black Holes and Naked Singularities
       in Low Energy Limit of String Gravity with Modulus Field}

\author{S. Alexeyev${}^1$\ftnote{1}{alexeyev@grg2.phys.msu.su},
        S.Mignemi${}^{2,3}$\ftnote{2}{mignemi@ca.infn.it}}

\address{${}^1$ Sternberg Astronomical Institute,
                Moscow State University, \\
                Universitetskii Prospect, 13,
                Moscow 119899, Russia; \\
         ${}^2$ Dipartimento di Matematica, Universit\`a  di Cagliari, \\
                viale Merello 92, 09123, Cagliari, Italy; \\
         ${}^3$ INFN, Sezione di Cagliari.}

\begin{abstract}
We show  that the black hole solutions of  the effective string theory
action, where one-loop effects that couple the moduli to gravity via a
\GB  term are  taken  into account, admit  primary  scalar hair.  The
requirement of absence of  naked  singularities imposes an upper bound
on the scalar charges.
\end{abstract}


\section{Introduction}

General Relativity describes very well gravity at the classical level,
but a quantum theory of  gravity  requires the introduction of a  more
general framework. One of  the  most promising candidates is presently
string  theory.  This  theory  is believed to change  drastically  the
short-range behavior of classical gravity, but also some of its global
properties can be modified, such as  black  hole  thermodynamics.  For
example, the study of the \bh solutions of effective low-energy theory
has  shown  that,  due  to  the  presence  of  non-minimal  couplings,
non-trivial  scalar  hair  can  arise  \cite{GHS},  in  contrast  with
classical general  relativity,  where no-hair theorems \cite{Bek} rule
out this possibility.

The  effects  of string theory on gravitational  physics  are  usually
investigated by  means  of  effective  field  theory actions, obtained
through a  perturbative expansion in the  string tension $\a$.  At the
tree level, the  effective  action of  the  heterotic (but also  other
types of)  string contains a coupling of the  dilaton with gravity via
the \GB term. The \bh  solutions  of this model have been  extensively
studied  in  the  literature,  both  in  a perturbative \cite{MS}  and
numerical \cite{Al,KM} approach. It turns  out  that  the model admits
\af \bh solutions with  non-trivial  dilatonic hair. The scalar charge
is not an independent parameter, but  is a function of the mass of the
black hole,  and is therefore  an example of secondary hair \cite{Co}.
Also the thermodynamics is different from that of \Sch black holes. In
particular, it was shown that the theory predicts a lower bound on the
mass of  \GB black holes  \cite{Al,KM}, which corresponds to the state
of   highest   (but  finite)  temperature  and  lowest  entropy.   The
configuration of minimal mass should be  identified  with  the  ground
state of the Hawking evaporation process.

In  order to  build  realistic models, one  should  however take  into
account that  in string theory other scalar fields  are present in the
spectrum in addition to the dilaton, as for example the  moduli, which
originate  from   the   compactification   of  the  higher-dimensional
spacetime. These also couple  to  gravity through one-loop effects. At
leading order the coupling term is proportional to the logarithm  of a
Dedekind $\y$-function of the moduli,  which  multiplies  the \GB term
\cite{An}. The effect  of the non-minimal  coupling of the  moduli  to
gravity in  a cosmological context  has been studied in several papers
\cite{EM} and it has been shown that in some  cases  it  may  lead  to
models  without  initial  singularities,  but  to   our  knowledge  no
investigation has been  devoted till now  to its implications  on  \bh
physics.

On the other hand,  it is well known that in effective  string actions
the  electromagnetic  field  exhibits  a non-minimal coupling  to  the
dilaton and the moduli similar to  that of the \GB term \cite{Ka}. The
\bh  solutions  have been  thoroughly  studied in  this  case: if  one
neglects the moduli, one obtains exact magnetically charged solutions,
with secondary scalar hair, the scalar charge being a function  of the
mass and the  magnetic charge \cite{GHS}.  If one instead  takes  into
account  also  one modulus,  the  general solution  can  no longer  be
written in  analytic form, but it can nevertheless  be shown to depend
on three parameters  \cite{Mi}: thus in  this case a  new  independent
parameter arises,  besides the mass and the charge,  and one may speak
of a primary scalar hair.

In this paper, we investigate if a similar phenomenon can occur in the
purely gravitational sector. Since we are mainly interested in showing
the existence of primary scalar hair in the  scalar-gravity sector, we
consider a simplified model with a dilaton and a unique  modulus which
couples  exponentially with the  \GB  term.  We  study it  both  in  a
perturbative and numerical  setting  using the techniques developed in
Ref. \cite{MS}  and \cite{Al}, respectively.  We find that in fact the
qualitative features  are similar to  the case of Maxwell coupling. We
also find  that an  upper limit must be imposed  on the scalar charges
for  given  mass, in  order  to  avoid  naked  singularities.  This is
reminiscent   of   the   extremality   bounds   in   the   multiscalar
Einstein-Maxwell case \cite{MW}.

The structure of our  paper is the following. In section 2  we present
the perturbative  solution and discuss its thermodynamical properties.
In section 3 we describe the numerical solution and the  occurrence of
an upper bound for the scalar charges. Section 4 contains a discussion
and the main conclusions.

\section{Perturbative Solution}

The bosonic sector of the effective action for the heterotic string in
absence of Yang-Mills and axionic fields, is given at leading order in
$\alp$ by
\be
I_{eff}={1\over 16\pi}\int d^4 x \sqrt{-g}\left[\ri-2(\nabla\Phi)^2
-2(\nabla \Sigma)^2+\alpha(\ef+\d\es)\gb\right]
\ee
where  $\alpha\equiv\alp  /8$, $\d$ is a coupling  constant  of  order
unity, $\F$ is the dilaton, $\S$ is a modulus, whose coupling with the
\GB term, $\gb\equiv\ri_{mnpq}\ri^{mnpq}-4\ri_{mn}\ri^{mn}+\ri^2$  has
been taken for simplicity to be of exponential form. The \fe equations
can be written as
\bea\label{fe}
&&G_\mn=T_\mn^{(\F)}+T_\mn^{(\S)},\nn\\
&&\nabla^2\F={\alpha\over 2}\ef\gb,\qquad\qquad\nabla^2\S={\a\d\over 2}\es\gb,
\eea
where
\bea
\le{T_\mn^{(\F)}=4\alpha\ef\Big[4\ri_{p(m}\nabla_{n)}\nabla_p\Phi-
2\ri_{mn}\nabla_p\nabla_p\Phi-
\ri\nabla_m\nabla_n\Phi-2\ri_{qmnp}\nabla_p\nabla_q\Phi\Big]\nn}\\
&&-8\alpha\ef\Big[ 4\ri_{p(m}\nabla_{n)}\Phi\nabla_p\Phi-2\ri_{mn}\nabla_p\Phi
\nabla_p\Phi-\ri\nabla_m\Phi\nabla_n\Phi-2\ri_{qmnp}
\nabla_p\Phi\nabla_q\Phi\Big]\nn\\
&&+2\nabla_m\Phi\nabla_n\Phi-g_{mn}\nabla_m\F\nabla_n\F,
\eea
and   an   alougous   expression   for   $T_\mn^{(\S)}$.   The   total
energy-momentum is conserved but, as noticed in \cite{KM} for the case
of a single  scalar, its time  component, corresponding to  the  total
energy, is  not positive definite, due to the  contribution of the \GB
term, leaving  room for the possibility of a  violation of the no-hair
conjecture.

We look for \ssy solutions,  with  scalars  $\F=\F(r)$, $\S=\S(r)$.
A generic \ssy metric can be written as
\be\label{metric}
\ds-\D(r)dt^2+{\s^2(r)\over\D(r)}dr^2+R^2(r)d\O^2.
\ee
Instead of substituting the ansatz (\ref{metric})
into the field equations (\ref{fe}),
it is easier to substitute it directly into the action,
and then vary with respect
to the fields $\F$, $\S$, $\D$, $\s$ and $R$.

The action reads:

\bea
\le{I={1\over 16\pi}\int d^4 x {2\over\s}\Big[\D'RR'+\D R'^2+\s^2
-R^2\D(\F'^2+\S'^2)\nn}\\
&&-2\a\efs'\D'\left(1-\s^\mt\D R'^2\right)\Big]
\eea

The \fe can then be written in the form:
\bea\label{feq}
\le{(\s^\mo\D R^2\F')' = -2\a\ef[(\s^\mo\D'(1-\s^\mt\D R'^2)]',}\nn\\
\le{(\s^\mo\D R^2\S')' = -2\a\d\es[(\s^\mo\D'(1-\s^\mt\D R'^2)]',\nn}\\
\le{RR''-\s^\mo\s'RR' = 2\a\s^\mo V'[2\s^\mo\D R'R''-(1-3\s^\mt\D R'^2)]
\nonumber}\\
&&- 2\a V''\s^\mo(1-\s^\mt\D R'^2)- R^2\fsq ,\\
\le{\D'RR' + \D R'^2 -\s^{-2}= - 2\a V'\D'(1 - 3\s^\mt\D R'^2)+\D R^2\fsq,}
\nonumber\\
\le{\D''R + 2\D'R' + 2\D R'' -\s^\mo\s'(\D'R+2\D R')\nonumber}\\
&&= 4\a V'(\s^{-3}\D\D'R')'+ 4\a V''\s^{-3}\D\D'R'-2R\D\fsq,\nonumber
\eea
where $V=\efs$ and  a prime denotes  derivative with respect  to  $r$.
Only four of the previous equations are independent.

In order to  find  approximate solutions  to  the field equations,  we
adopt the approach of Ref. \cite{MS} and expand the fields  around the
background constituted  by  the  Schwarzschild  metric  with vanishing
scalar fields, which is of course  a solution for $\a=0$.
Our expansion is in the
parameter $\alpha/m^2$,  $m$ being the  mass of the \bg \Sch solution.
Since $\alpha$ is believed  to be of order unity in Planck  units, the
expansion   is   valid   for   large   $m$,   in   the   region  where
$\alpha\gb\ll\ri$, i. e.  for  $r^3\gg\alpha m$. For macroscopic black
holes  ($m\gg 1$)  this  condition is always  satisfied,  except in  a
neighborhood of the singularity, well  inside  the  horizon (region of
strong curvature). In particular,  the  approximation is valid for the
discussion  of  the  asymptotic  properties  of  the  fields,  and the
questions concerning the  scalar  hair. Near the physical singularity,
however,  the  higher  order  corrections  to   the  effective  string
lagrangian become important and  the  perturbation theory is no longer
reliable.  We  shall  however  discuss  this  regime  using  numerical
techniques in the next section.

At this point, it must be observed that the ansatz  (\ref{metric}) for
the metric is too general and still leaves the possibility of a choice
of gauge. In order to perform the perturbative  calculations, the most
convenient choice \cite{MS} is to impose $\s\equiv1$, i.e.
\be\label{GHS}
\ds-\D(r)dt^2+{1\over\D(r)}dr^2+R^2(r)d\O^2.
\ee
This gauge was also used
for finding exact charged \bh solutions in effective string theory
\cite{GHS}.

We expand the fields as follows:
\bea\label{exp}
\D & = & \D_0(1+\a\q_1+\a^2\q_2+\a^3\q_3+\dots),\nn\\
R & = & r+\a\r_1+\a^2\r_2+\a^3\r_3+\dots,\nonumber\\
\Phi & = & \a\Phi_1+\a^2\Phi_2+\a^3\Phi_3+\dots,\nonumber\\
\Sigma & = & \S_0+\a\S_1 + \a^2\S_2 + \a^3\S_3+\dots,
\eea
where $\D_0=(1-2m/r)$. We have  normalized  $\F$ such that $\F\to0$ at
infinity. This is  always possible, by rescaling the coupling constant
$\a$ (this means  that our expansion is actually in $\a\ex^{-2\F_0}$).
However, it  is not possible to  rescale independently also  $\S$, and
hence we take $\S\to\S_0$ at infinity. The parameters  $\d$ and $\S_0$
will always appear in the combination $Z=\d\ex^{-2\S_0}$.

Substituting the expansion  (\ref{exp})  into the \fe (\ref{feq}), one
obtains at first order
\be
\lap\fo']'={24m^2\over r^4},\qquad\lap\so']'={24m^2Z\over r^4}.
\ee
With the previous \bc, requiring regularity at the horizon $r=2m$, the
scalar fields are uniquely determined at first order:
\be
\fo = {\so\over Z} =
-{1\over m}\left( {1\over r}+{m\over r^2}+{4m^2\over 3r^3}
\right),
\ee
The equations  for the metric fields are  given at  the same order  by
\be
\ro ''=0,\qquad\qquad\lapp\po]'=-{2m\over r^2}\ro.
\ee

We impose  the boundary conditions  that $\ro\to$ const, $\po\to 0$ at
infinity.  We are  still  free to choose  the  boundary conditions  at
$r=2m$. Changing  the \bc at  $r=2m$ yields a reparametrization of the
solutions, but no  change in their physical properties: in particular,
the relations between the physical quantities,  like mass, temperature
and  entropy,  are  independent  of  the   parametrization.  The  most
convenient  choice is  to  require that the  $\r_i$  and $\psi_i$  are
regular at $r=2m$.  This  is  equivalent to fix the  location  of  the
horizon at $r=2m$.  With  these boundary conditions, $\ro=0$, $\po=0$.
This choice greatly simplifies the higher order calculations.

We can now  evaluate  the second  order  corrections. With the  stated
boundary conditions, the equations for the  second order perturbations
give
\bea
\le{\lap\ft']'=-{48m^2\over r^4}\fo\nn,}\\
\le{\lap\st']'=-{48Zm^2\over r^4}\so\nn,}\\
\le{\rt''=-r(\fo'^2+\so'^2)+{8m\over r^2}(\fo''+Z\so''),}\\
\le{\lapp\pt]'=-(r-2m)\rt ''-2{r-m\over r}\rt '-{2m\over r^2}\rt}\nn\\
&&+8m{r-2m\over r^2}(\fo''+Z\so'')-16m{r-3m\over r^3}(\fo'+Z\so'),\nn
\eea
whose solution is
\bea
&&\Phi_2={\S_2\over Z^2}=-{1\over m^3}\left(
 {73\over60r}+{73m\over60r^2}+{73m^2\over45r^3}+{73m^3\over30r^4}+
 {112m^4\over75r^5}+{8m^5\over9r^6}\right),\nonumber\\
&&\rho_2=-{1 + Z^2\over m^2}\left(
 {1\over2r}+{2m\over3r^2}+{7m^2\over3r^3}+{16m^3\over5r^4}+{24m^4\over5r^5}
 \right),\nonumber\\
&&\psi_2=-{ 1 + Z^2\over m^3 }\left(
 {1\over6r}+{m\over3r^2}+{4m^2\over3r^3}-{14m^3\over3r^4}-{136m^4\over15r^5}
 -{272m^5\over15r^6}\right).\nonumber
\eea
Up  to  this  order, $\F$ and  $\S$  are  proportional  in this gauge.
However, in order  to  clarify the structure of  the  solutions, it is
useful to  go to the  next order,  even if such  corrections would  of
course be modified by taking into account terms of order $\a^2$ in the
action. A long but straightforward calculation gives
\bea
\rho_3=-{73\,( 1 + Z^3)\over {60\,{m^4}\,r}}+\orq,\qquad
\psi_3=-{73\,( 1 + Z^3)\over {180\,{m^5}\,r}}+\orq,\nonumber
\eea
\bea
\F_3=- {16480 + 3969\,Z^2\over {7560\,{m^5}\,r}}+\orq,\qquad
\S_3 = - {Z\, (3969 + 16480\,Z^2) \over {7560\,{m^5}\,r}}+\orq.\nonumber
\eea
From these  results appears that  the metric functions are expanded in
terms  of $\a^k(1+Z)^k$,  while  the scalar fields  depend  in a  more
involved way from the parameters. Moreover, the functional dependences
of the dilaton and the modulus on $r$ are different if $Z\ne1$.

The perturbative solutions have the following properties: a horizon is
present at $r=2m$, while a singularity  is located at the zero of $R$;
the evaluation  of this zero is however outside  the range of validity
of our  approximation. It can  be expected nevertheless that for small
values of the mass, or  great  values  of  $Z$, the zero can occur for
$r>2m$, leading  to the presence  of naked singularities. This will be
confirmed by the numerical results of next section.

The mass  $M$ of the  black hole  can be deduced  from the  asymptotic
behavior of the metric function $\D$ and is given by
\be\label{mass}
M=m\left(1+{1\over12}{\alpha^2(1+Z^2)\over m^4}
    +{73\over360}{\alpha^3(1+Z^3)\over m^6}\right).
\ee
Its value  is greater than  that of  the \Sch \bh  with equal  radius.
Analogously, from  the asymptotic behaviour of  $\F$ and $\S$  one can
deduce the scalar charges $D_\F$  and  $D_\S$, which, in terms of  the
mass $M$, turn out to be
\bea
&&D_\F={\a\over M}\left(1+{73\over60}{\alpha\over M^2}
    +{17110+4599Z^2\over7560}{\alpha^2\over M^4}\right),\nn\\
&&D_\S={\a Z\over M}\left(1+{73\over60}{\alpha Z\over M^2}
    +{4599+17110Z^2\over7560}{\alpha^2\over M^4}\right).
\eea

It is clear that, contrary to  the case in which only one scalar field
is present  \cite{MS}, the scalar  charges are no longer function only
of the mass of  the black hole, but depend also on  another parameter,
which we identify with $Z$. Hence, in analogy with the dilaton-modulus
gravity non-minimally  coupled to the electromagnetic field \cite{Mi},
also in  this case  a primary scalar hair is  present in the solution.
This gives  an example of primary scalar hair  in pure gravity models.
We notice that, at leading order, $Z\sim D_\S/D_\F$.

The temperature  of the \bh  can be  defined as usual  as the  inverse
periodicity of the time coordinate which renders regular the Euclidean
section of the metric. This is given by
\cite{Yo}
\be\label{temp}
T={1\over4\pi\sqrt{g_{00}g_{11}}}{dg_{00}\over dr}\Bigl\arrowvert_
{\rm hor}\, ,\nn
\ee
which yields, at order $\alpha^3$:
\be
T={1\over8\pi m}\big(1+\a^2\pt(2m)+\a^3\q_3(2m)\big).\nn
\ee
Taking into account (\ref{mass}), a straightforward calculation leads to
\be
T={1\over8\pi M}\left(1+{73\over120}{\alpha^2(1+Z^2)\over M^4}
    +{12511\over7560}{\alpha^3(1+Z^3)\over M^6}\right).\nn
\ee
The temperature is higher than that of a \Sch \bh  of  equal mass, but
is still a monotonically decreasing function of the mass.

The entropy $S$  can  be defined by means  of  the Euclidean formalism
\cite{Ha} as
\be\label{entropy}
S=\beta{\partial I_E\over\partial\beta}-I_E,
\ee
where $\beta$ is the inverse temperature and $I_E$ the Euclidean action
\bea\label{eaction}
I_E&=&-{1\over 16\pi}\int_M[\ri-2(\nabla\Phi)^2-2(\nabla\S)^2+
\alpha\ (\ef+\d\es)\gb)]dV\nn\\
&&-{1\over 8\pi}\int_{\partial M}(K-K_0)d\Sigma
\eea
where $K$ is the exterior curvature. A lenghty calculation gives
\be
S=4\pi M^2\left(1+{73\over120}{\alpha^2(1+Z^2)\over M^4}
    +{12511\over15120}{\alpha^3(1+Z^3)\over M^6}\right).\nn
\ee

In the range of validity of  our  approximation,  the  thermodynamical
quantities of course do not differ much from  their background values,
except that now they depend on the further parameter $Z$.  They behave
differently only for $M\ll\a$, where however  the approximation breaks
down.  It  is  interesting  to  notice  that  all  the thermodynamical
quantities are expanded in terms of $\a^k(1+Z^k)$.

\section{Numerical Solution and Naked Singularity}

As discussed previously, the perturbative solution is not valid in the
whole domain of definition  of the solution. Since we are not  able to
find the exact analytical solution of the field equations, we  use the
numerical approach described  in  Ref. \cite{Al}. For this calculation
it is more convenient  to choose a gauge in which the  metric function
$R$ is identified with the radial coordinate $r$, i.e.
\begin{eqnarray}\label{e2.1}
ds^2 = -\D dt^2 + \frac{\sigma^2 }{\D } dR^2 + R^2 d\O^2 \nonumber,
\end{eqnarray}
where $\D=\D(R)$,  $\s=\s(R)$.  Comparison  with (\ref{metric}) yields
$\s(R)=dr/dR$. Of course, the physical quantities do not depend on the
choice of  gauge. However, in order to make  the comparison of results
easier, we give the perturbative expansions in the new coordinates:
\bea
\le{\D(R)=(1-2m/R)(1+\a^2\q_2(R)+\a^3\q_3(R))+(2m/R^2)(\a^2\rt(R)+\a^3\r_3(R))
+\dots\nn}\\
\le{\s(R)=1-\a^2\r_2'(R)-\a^3\r_3'(R)+\dots\nn}\\
\le{\F(R)=\F_0+\a\F_1(R)+\a^2\F_2(R)+\a^3(\F_3(R)-\F_1'(R)\r_2(R))+\dots\nn}\\
\le{\S(R)=\S_0+\a\S_1(R)+\a^2\S_2(R)+\a^3(\S_3(R)-\S_1'(R)\r_2(R))+\dots\nn}
\eea
where  use  has been  made  of the  condition  $\q_1=\r_1=0$, and  the
functions $\r_i$, $\q_i$\, $\F_i$ and $\S_i$ are those obtained above,
evaluated at $r=R$.  In  particular, the  metric  function are, up  to
order $\a^2$,
\bea
\D&=&1-{2m\over R}-{\a^2(1+Z^2)\over m^4}\left({m\over6R}+{5m^3\over3R^3}
-{6m^4\over R^4}+{74m^5\over15R^5}+{32m^6\over5R^6}+{688m^7\over15R^7}
\right)\nn\\
\s&=&1-{\a^2(1+Z^2)\over m^4}\left({m^2\over2R^2}+{4m^3\over3R^3}
+{7m^4\over R^4}+{64m^5\over5R^5}+{24m^6\over R^6}\right)\nn
\eea

In the  parametrization  (\ref{e2.1})  the Einstein-Lagrange equations
can  be written  in a  matrix  form (we  set all  the string  coupling
constants to be equal to one for simplicity)
\begin{eqnarray}\label{syst1}
a_{i1} \Delta'' + a_{i2} \sigma'
+ a_{i3} \Phi'' + a_{i4} \Sigma'' = b_i,
\end{eqnarray}
where $i=1, \ldots,4$  and the entries  of the matrices  $a_{ij}$  and
$b_i$ are
\begin{eqnarray}
a_{11} & = & 0                                         \nonumber \\
a_{12} & = & - \sigma^2 R
       + 4 (\sigma^2 - 3 \Delta)
       ( e^{-2\Phi} \Phi'
       + e^{-2\Sigma} \Sigma')                         \nonumber \\
a_{13} & = & 4 \sigma (\Delta-\sigma^2) e^{-2\Phi}     \nonumber \\
a_{14} & = & 4 \sigma (\Delta-\sigma^2) e^{-2\Sigma}   \nonumber \\
a_{21} & = & \sigma^3 R + 8 \Delta \sigma
             ( e^{-2\Phi} \Phi'+ e^{-2\Sigma} \Sigma') \nonumber \\
a_{22} & = & - \sigma^2 (\Delta' R + 2 \Delta)
        - 24 \Delta \Delta' ( e^{-2\Phi} \Phi'
        + e^{-2\Sigma} \Sigma')                        \nonumber \\
a_{23} & = & 8 \Delta \Delta' \sigma e^{-2\Phi}        \nonumber \\
a_{24} & = & 8 \Delta \Delta' \sigma e^{-2\Sigma}      \nonumber \\
a_{31} & = & 4 e^{-2\Phi} \sigma (\Delta-\sigma^2)     \nonumber \\
a_{32} & = & 2 \sigma^2 R^2 \Delta \Phi'
        -4 e^{-2\Phi} \Delta' (3 \Delta-\sigma^2)      \nonumber \\
a_{33} & = & -2 \sigma^3 R^2 \Delta                    \nonumber \\
a_{34} & = & 0                                         \nonumber \\
a_{41} & = & 4 e^{-2\Sigma} \sigma (\Delta-\sigma^2)   \nonumber \\
a_{42} & = & 2 \sigma^2 R^2 \Delta \Sigma'
        - 4 e^{-2\Sigma} \Delta' (3 \Delta-\sigma^2)   \nonumber \\
a_{43} & = & 0                                         \nonumber \\
a_{44} & = & -2 \sigma^3 R^2 \Delta                    \nonumber \\
\nonumber \\
b_1 & = & - \sigma^3 R^2 ( \Phi'^2 + \Sigma'^2)
       + 8 \sigma (\Delta-\sigma^2) (e^{-2\Phi} \Phi'^2
       + e^{-2\Sigma} \Sigma'^2 )                     \nonumber \\
b_2 & = & -2 \sigma^3 (\Delta'+\Delta R ( \Phi'^2 + \Sigma'^2))
       +16 \Delta \Delta' \sigma (e^{-2\Phi} \Phi'^2
       + e^{-2\Sigma} \Sigma'^2 ) \nonumber \\
       & - & 8 \Delta'^2 \sigma (e^{-2\Phi} \Phi'
       + e^{-2\Sigma} \Sigma')                        \nonumber \\
b_3 & = & 2 \sigma^3 R \Phi' (\Delta' R + 2 \Delta)
       -4 e^{-2\Phi} \Delta'^2 \sigma                 \nonumber \\
b_4 & = & 2 \sigma^3 R \Sigma' (\Delta' R + 2 \Delta)
       -4 e^{-2\Sigma} \Delta'^2  \sigma              \nonumber
\end{eqnarray}

The last (constraint) equation is
\begin{eqnarray}
&& \sigma^2 \biggl[ \Delta R^2 (\Phi'^2 + \Sigma'^2) + \sigma^2
                   - \Delta' R - \Delta \biggr] \nonumber \\
& + & 4 \biggl[ \lambda_\Phi e^{-2 \Phi} \Phi'
    + \lambda_\Sigma q e^{-2 q \Sigma} \Sigma' \biggr]
    \Delta' (\sigma^2 - 3 \Delta) = 0 \nonumber
\end{eqnarray}

The numerical integration  is performed as  follows: we start  from  a
neighborood of the horizon and  integrate  towards  infinity. The mass
and charges of  the solution are  then evaluated from  the  asymptotic
behaviour of the metric functions and  the  integration  is  performed
again backwards. More technical details on the numerical procedure can
be found in \cite{Al}.

The behaviour of  the solution outside  the horizon agrees  with  that
obtained by perturbative  methods in the previous section (see Figures
1-4) and  is similar to  that of the single-scalar solutions discussed
in \cite{MS,Al,KM}. When $D_\Sigma$ vanishes, we recover the solutions
described in  Ref. \cite{Al}. In particular,  one must impose  a lower
bound  on the  \bh mass  in  order to  avoid the  occurrence of  naked
singularities. For non-zero value of $D_\Sigma$  the situation changes
significantly.  Taking  $M$  and  $D_\F$  fixed,  for small values  of
$D_\Sigma$ the behavior of the solution does not differ much  from the
case of a single  scalar field, except that the position of  the inner
black hole singularity slowly moves up. This situation is shown by the
solid  lines  in  Figs.1-4,  where  the  dependence  of  the functions
$\Delta$, $\sigma$, $e^{-2\Phi}$ and $e^{-2\Sigma}$ against the radial
coordinate $r$ is plotted. When  $D_\Sigma$  reaches  a critical value
${D_\Sigma}_{crit}$ the positions of the  singularity  $R_s$  and  the
horizon $R_h$  coincide. The dependence of ${D_\Sigma}_{crit}$ against
black  hole  mass  is  represented  on  Figure  5.  When  $D_\Sigma  >
{D_\Sigma}_{crit}$ a  naked  singularity  appears.  This  situation is
shown  by  dashed  lines in Figs.  1-4.  This  naked  singularity is a
continuation of the  black  hole inner  singularity  and has the  same
nature. Of  course, since the system is symmetric  for the exchange of
$\F$ and  $\S$, an analogous behavior is expected  when $\F$ is varied
and $\S$ held fixed. The  dependence  of  ${D_\Sigma}_{crit}$ from the
black hole  mass $M$ is  approximately linear. Hence, according to the
cosmic censorship conjecture, the mass of the \bh gives an upper limit
for the modulus/dilatonic field charge.

\begin{figure}
\epsfxsize=0.7\hsize
\epsfysize=0.4\hsize
\centerline{\epsfbox{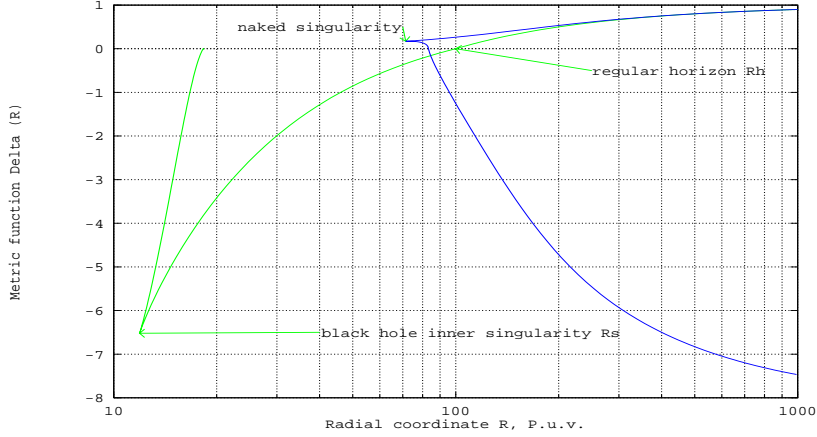}}
\caption{Dependence of the metric function $\Delta$ against the radial
coordinate   $R$,    Planck    unit   values,   when   $D_\Sigma   \ll
{D_\Sigma}_{crit}$ (thin  line)  and  $D_\Sigma \gg {D_\Sigma}_{crit}$
(thick line).}
\end{figure}

\begin{figure}
\epsfxsize=0.7\hsize
\epsfysize=0.4\hsize
\centerline{\epsfbox{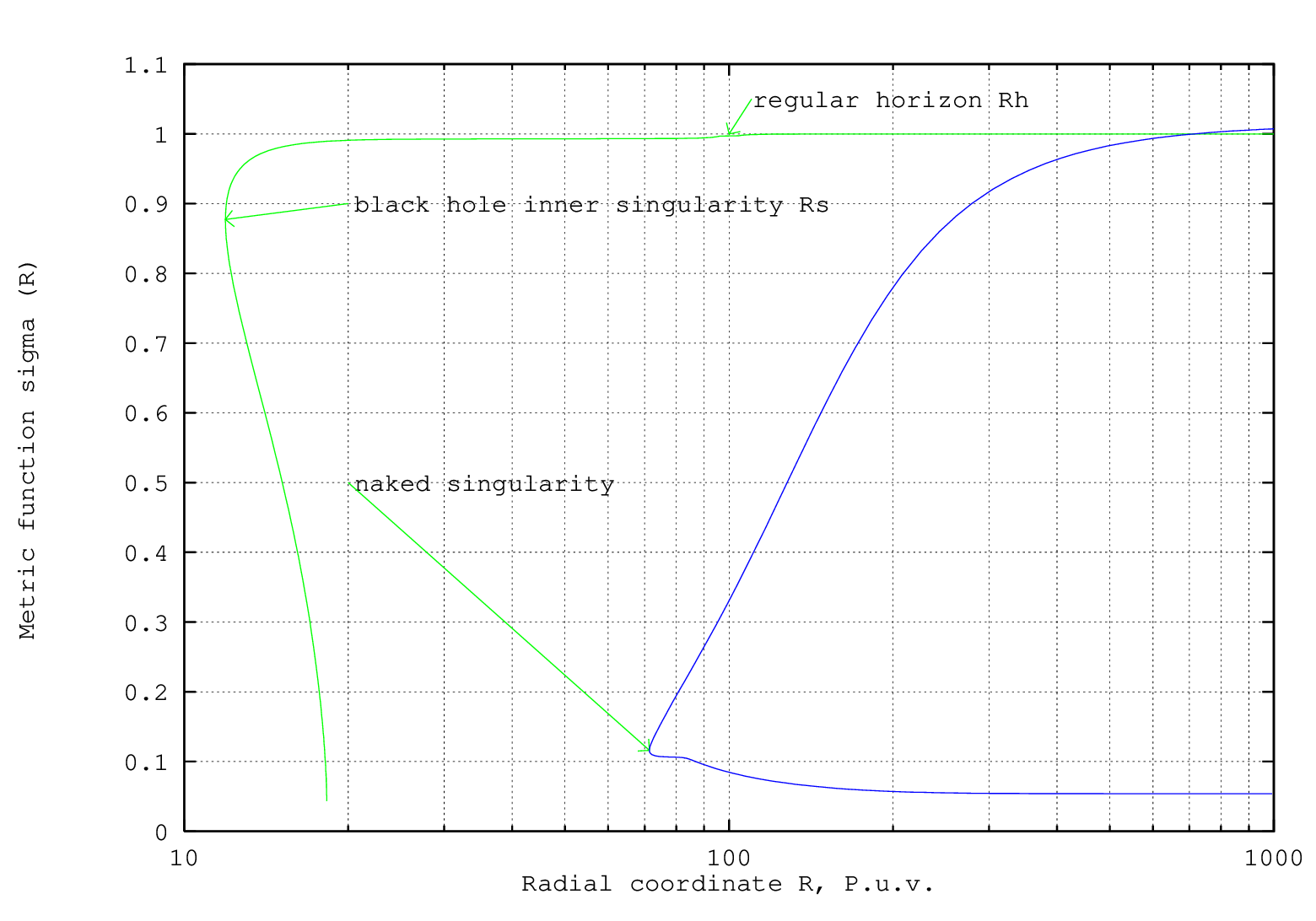}}
\caption{Dependence of the metric function $\sigma$ against the radial
coordinate   $R$,    Planck    unit   values,   when   $D_\Sigma   \ll
{D_\Sigma}_{crit}$ (thin  line)  and  $D_\Sigma \gg {D_\Sigma}_{crit}$
(thick line)}
\end{figure}

\begin{figure}
\epsfxsize=0.7\hsize
\epsfysize=0.4\hsize
\centerline{\epsfbox{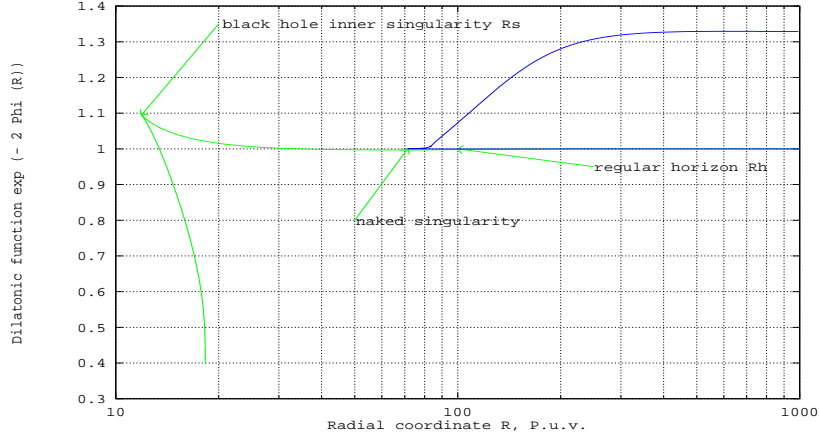}}
\caption{Dependence of  the  dilatonic  function $e^{-2 \Phi}$ against
the  radial  coordinate $R$, Planck unit values,  when  $D_\Sigma  \ll
{D_\Sigma}_{crit}$ (thin  line)  and  $D_\Sigma \gg {D_\Sigma}_{crit}$
(thick line)}
\end{figure}

\begin{figure}
\epsfxsize=0.7\hsize
\epsfysize=0.4\hsize
\centerline{\epsfbox{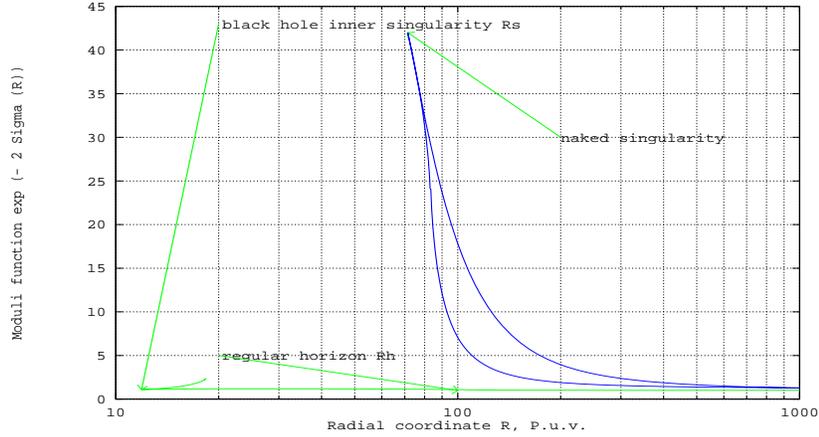}}
\caption{Dependence of  the  modulus  function $e^{-2 \Sigma}$ against
the radial coordinate $R$ when $D_\Sigma  \ll {D_\Sigma}_{crit}$ (thin
line) and $D_\Sigma \gg {D_\Sigma}_{crit}$ (thick line)}
\end{figure}
From a  mathematical point of  view the appearance of this singularity
is a  consequence of the vanishing of the  second factor (in brackets)
of the determinant $D_{main}$ of the system (\ref{syst1})
\begin{eqnarray}
D_{main} & = & \Delta^2 \biggl[ A \Delta^2
+ B \Delta + C \biggr] , \qquad
\mbox{where} \nonumber \\
A & = & 64 r^2 \sigma^5 \biggl[
        4 r^2 \sigma^2 e^{-4\Phi} \Phi'^2
     +  8 r^2 \sigma^2 e^{-2\Phi} \Phi' e^{-2\Sigma} \Sigma'
     +  4 r^2 \sigma^2 e^{-4\Sigma} \Sigma'^2 \nonumber \\
   & - &  \sigma^2 e^{-4\Phi}
     -  \sigma^2 e^{-4\Sigma}
     +  12 e^{-6\Phi} \Phi' \Delta'
     +  12 e^{-4\Phi} e^{-2\Sigma} \Sigma' \Delta' \nonumber \\
   & + & 12 e^{-2\Phi} \Phi' e^{-4\Sigma} \Delta'
     +  12 e^{-6\Sigma} \Sigma' \Delta' \biggr] \nonumber \\
B & = &  32 r^2 \sigma^6 \biggl[
         r^3 \sigma^3 e^{-2\Phi} \Phi'
      +  r^3 \sigma^3 e^{-2\Sigma} \Sigma'
      +  r^3 \sigma^2 e^{-2\Phi} \Phi'
      +  r^3 \sigma^2 e^{-2\Sigma} \Sigma' \nonumber \\
    & + & 2 r \sigma e^{-4\Phi} \Delta'
      +  2 r \sigma e^{-4\Sigma} \Delta'
      +  2 r e^{-4\Phi} \Delta'
      +  2 r e^{-4\Sigma} \Delta'
      +  4 \sigma^3 e^{-4\Phi}  \nonumber \\
    & + & 4 \sigma^3 e^{-4\Sigma}
      -  16 \sigma e^{-6\Phi} \Phi' \Delta'
      -  16 \sigma e^{-4\Phi} e^{-2\Sigma} \Sigma' \Delta' \nonumber \\
    & - & 16 \sigma e^{-2\Phi} \Phi' e^{-4\Sigma} \Delta'
      -  16 \sigma e^{-6\Sigma} \Sigma' \Delta'\biggr] \nonumber \\
C & = &  4 r^2 \sigma^8 \biggl[
          r^4 \sigma^2
       -  16 r \sigma e^{-4\Phi} \Delta'
       -  16 r \sigma e^{-4\Sigma} \Delta'
       -  16 r e^{-4\Phi} \Delta' \nonumber \\
     & - & 16 r e^{-4\Sigma} \Delta'
       -  16 \sigma^3 e^{-4\Phi}
       -  16 \sigma^3 e^{-4\Sigma}
       -  64 \sigma e^{-6\Phi} \Phi' \Delta' \nonumber \\
     & - & 64 \sigma e^{-4\Phi} e^{-2\Sigma} \Sigma' \Delta'
       -  64 \sigma e^{-2\Phi} \Phi' e^{-4\Sigma} \Delta'
       -  64 \sigma e^{-6\Sigma} \Sigma' \Delta' \biggr] \nonumber
\end{eqnarray}
A  naked  singularity  occurs  when  the  factor  in  the  bracket  of
$D_{main}$ vanishes  before  the  metric  function  $\Delta$. Figure 6
represents the  dependence of the position  of the zero  of $D_{main}$
against the parameter $Z$.

\begin{figure}
\epsfxsize=0.7\hsize
\epsfysize=0.4\hsize
\centerline{\epsfbox{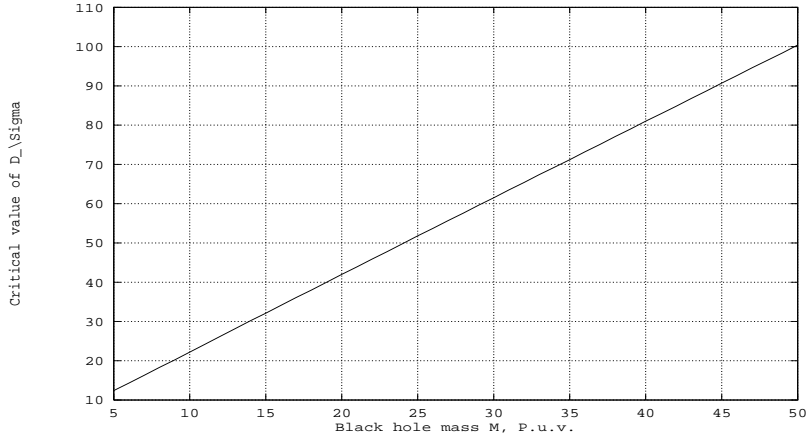}}
\caption{The    dependence    of    the   critical   modulus    charge
${D_\Sigma}_{crit}$ against the black hole mass  $M$  in  Planck  unit
values, for $D_\Phi=1$.}
\end{figure}

\begin{figure}
\epsfxsize=0.7\hsize
\epsfysize=0.4\hsize
\centerline{\epsfbox{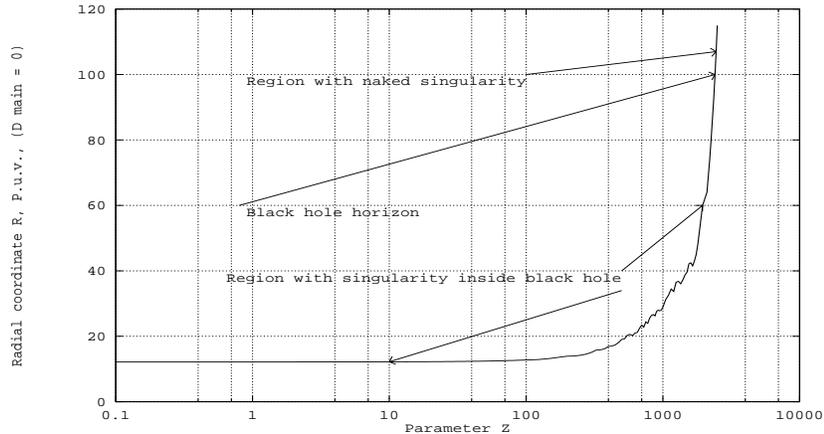}}
\caption{The dependence of  the value of $R$ for which
$D_{main}=0$, against the parameter $Z=D_\Sigma/D_\Phi$.}
\end{figure}

The thermodynamical  parameters can evaluated numerically and compared
with the perturbative results. This is interesting especially in order
tounderstand  their  behaviour for small mass, where the  perturbative
approach fails. The temperature can be obtained from (\ref{temp}) in a
straightforward  way.  For   the   calculation  of  the  entropy,  eq.
(\ref{entropy}) has been used. In  particular,  the  Euclidian  action
$I_E$ was evaluated by adding its definition as an additional equation
to the main system (\ref{syst1}).

The numerical evaluation of the black hole temperature $T$ and entropy
$S$ agrees with the perturbative results for great $M$, see  the Table
1. A similar agreement holds for the entropy.

\begin{table}
\caption{The dependence of  the black hole temperature $T$ against the
mass  $M$  and  the  parameter  $Z  \approx  D_\Sigma/D_\Phi$  for the
numerical and  perturbative  solutions.  The  agreement  is better for
great $M$  or small  $Z$, in accordance with the  limit of validity of
the perturbative approach.}
\vskip3mm
\begin{center}
\begin{tabular}{|l|l|l|l|}\hline
$M$   & $Z$  & $T$ numerical             & $T$ perturbative          \\ \hline
 4.0  &  0.1 & $1.032079 \ * \ 10^{-02}$ & $9.975080 \ * \ 10^{-03}$ \\
      &  1.0 & $1.041815 \ * \ 10^{-02}$ & $1.000250 \ * \ 10^{-02}$ \\ \hline
 5.0  &  0.1 & $8.094905 \ * \ 10^{-03}$ & $7.966414 \ * \ 10^{-03}$ \\
      &  1.0 & $8.169681 \ * \ 10^{-03}$ & $7.974924 \ * \ 10^{-03}$ \\
      & 10.0 & $1.007615 \ * \ 10^{-02}$ & $9.583720 \ * \ 10^{-03}$ \\ \hline
10.0  &  0.1 & $3.977655 \ * \ 10^{-03}$ & $3.979124 \ * \ 10^{-03}$ \\
      &  1.0 & $3.979633 \ * \ 10^{-03}$ & $3.979371 \ * \ 10^{-03}$ \\
      & 10.0 & $4.101194 \ * \ 10^{-03}$ & $4.009911 \ * \ 10^{-03}$ \\ \hline
20.0  &  0.1 & $1.986029 \ * \ 10^{-03}$ & $1.989444 \ * \ 10^{-03}$ \\
      &  1.0 & $1.985790 \ * \ 10^{-03}$ & $1.989452 \ * \ 10^{-03}$ \\
      & 10.0 & $1.989498 \ * \ 10^{-03}$ & $1.990252 \ * \ 10^{-03}$ \\ \hline
30.0  &  0.1 & $1.323841 \ * \ 10^{-03}$ & $1.326292 \ * \ 10^{-03}$ \\
      &  1.0 & $1.323821 \ * \ 10^{-03}$ & $1.326293 \ * \ 10^{-03}$ \\
      & 10.0 & $1.324732 \ * \ 10^{-03}$ & $1.326395 \ * \ 10^{-03}$ \\ \hline
40.0  &  0.1 & $9.931118 \ * \ 10^{-04}$ & $9.947186 \ * \ 10^{-04}$ \\
      &  1.0 & $9.931105 \ * \ 10^{-04}$ & $9.947189 \ * \ 10^{-04}$ \\
      & 10.0 & $9.933026 \ * \ 10^{-04}$ & $9.947426 \ * \ 10^{-04}$ \\ \hline
50.0  &  0.1 & $7.943050 \ * \ 10^{-04}$ & $7.957748 \ * \ 10^{-04}$ \\
      &  1.0 & $7.943053 \ * \ 10^{-04}$ & $7.957749 \ * \ 10^{-04}$ \\
      & 10.0 & $7.948299 \ * \ 10^{-04}$ & $7.957825 \ * \ 10^{-04}$ \\ \hline
\end{tabular}
\end{center}
\end{table}
\vskip3mm

Finally, we notice  that the thermodynamical parameters stay finite in
the extremal case. The  thermodynamics  is therefore analogous to that
studied in absence of modulus fields \cite{Al,KM}, except that now one
has one further independent parameter (the scalar charge).

\section{Discussion and conclusions}

We have shown perturbatively the occurrence of primary  scalar hair in
black  hole  solutions of  models  with  more  than  one  scalar field
non-minimally coupled to  gravity via the  \GB term. This  result  has
been checked numerically. From the numerical calculations also follows
that naked  singularities can appear for small values  of the mass (as
in  pure  dilaton-\GB models),  or  for  large  values  of  the scalar
charges. This is a novel feature  of the model under study, and can be
compared  with   a   similar   phenomenon  occurring  in  multi-scalar
Einstein-Maxwell  models  \cite{MW}.  In  that  case  some  analytical
relations for the  extremality  condition of  the  black holes can  be
obtained, while  in our case  this seems  not to be  possible. We  can
however conjecture that a relation of the same kind exists also in our
case.

\section*{Acknowledgments}

S.  M.  wishes  to  thank  Sternberg  Astronomical Institute for  kind
hospitality  during  the last  stages  of  this  work.  This  work was
partially supported by a coordinate research project of the University
of   Cagliari   and   by   ``Universities   of   Russia:   Fundamental
Investigations'' Program via grant No. 990777.


\section*{References}

\begin{thebibliography}{99}
\bibitem{GHS}
D. Garfinkle, G.T. Horowitz, and A. Strominger, {\it  Phys. Rev.} {\bf
D 43}, 3140 (1991).

\bibitem{Bek}
J.D. Bekenstein, {\it Phys. Rev.} {\bf D 5}, 1239 (1972);
J.E. Chase, {\it Comm. Math. Phys.} {\bf 19}, 276 (1970).

\bibitem{MS}
S. Mignemi and N.R. Stewart, {\it Phys. Rev.} {\bf D 47}, 5259 (1993).

\bibitem{Al}
S.O. Alexeyev and M.V. Pomazanov,  {\it  Phys. Rev.} {\bf D 55},  2110
(1997);
S.O. Alexeyev and M.V. Sazhin, {\it Gen. Relativ. and Grav.}  {\bf 8},
1187 (1998).

\bibitem{KM}
P. Kanti,  N.E. Mavromatos, J. Rizos,  K. Tamvakis and  E. Winstanley,
{\it Phys. Rev.} {\bf D 54}, 5049 (1996);
P. Kanti and K. Tamvakis, {\it Phys. Lett.} {\bf B 392}, 30 (1997);
T.  Torii,  H.  Yajima, and K. Maeda, {\it Phys. Rev.} {\bf D 55}, 739
(1997).

\bibitem{Co}
S. Coleman, J. Preskill, and F. Wilczek, {\it Nucl. Phys.} {\bf B380},
447 (1992).

\bibitem{An}
I.  Antoniadis, J.  Rizos,  and K. Tamvakis,  {\it  Nucl. Phys.}  {\bf
B415}, 497 (1994).

\bibitem{EM}
R. Easther and K. Maeda, {\it Phys. Rev.} {\bf D 54}, 7252 (1996);
P. Kanti,  J. Rizos, and  K. Tamvakis,  {\it Phys. Rev.}  {\bf D  59},
083512 (1999);
S. Alexeyev, A. Toporensky, V.  Ustiansky,  {\it  Class. Quant. Grav.}
{\bf 17}, 2243 (2000).

\bibitem{Ka}
V. Kaplunowsky, {\it Nucl. Phys.} {\bf B307}, 145 (1988).

\bibitem{Mi}
S. Mignemi, {\it Phys. Rev.} {\bf D 62}, 024014 (2000).

\bibitem{Yo}
J.W. York, {\it Phys. Rev.} {\bf D 31}, 775 (1985).

\bibitem{Ha}
S.W. Hawking, in {\it "General Relativity: an Einstein centenary
survey"}, eds. S.W. Hawking and W. Israel (Cambridge Un. Press 1979).

\bibitem{MW}
S. Mignemi and D.L. Wiltshire, in preparation.

\end{thebibliography}
\end{document}